\documentclass[useAMS,usegraphicx,usenatbib]{mn2e}
\usepackage{url}
\usepackage{amssymb}
\newcommand{\myr}{{q}}
\newcommand{\myin}{{i}}
\newcommand{\myout}{{o}}
\begin{document}

\title[Shock Induced Cold Fronts]{Cold Fronts by Merging of Shocks}

\author[Birnboim, Keshet \& Hernquist]{Yuval Birnboim$^{1}$, Uri
  Keshet$^{1}$\thanks{Einstein fellow} \& Lars Hernquist$^{1}$\\
$^{1}$Harvard-Smithsonian Center for Astrophysics, 60 Garden Street,
  Cambridge, MA 02138 USA}
\date{Accepted ---. Received ---; in original ---}

\pagerange{\pageref{firstpage}--\pageref{lastpage}} \pubyear{2010}

\bibliographystyle{mn2e}
\maketitle

\label{firstpage}

\begin{abstract}
Cold fronts (CFs) are found in most galaxy clusters, as well as in
some galaxies and groups of galaxies.  We propose that some CFs are
relics of merging between two shocks propagating in the same direction.
Such shock mergers
typically result in a quasi-spherical, factor $\approx 1.4-2.7$
 discontinuity in density and in temperature.  These CFs may be found as far out
as the virial shock, unlike what is expected in other CF formation
models.

As a demonstration of this effect, we use one dimensional simulations
of clusters and show that shock induced cold fronts form when
perturbations such as
explosions or mergers occur near the cluster's centre. Perturbations at
a cluster's core induce periodic
merging between the virial shock and outgoing secondary shocks.
These collisions yield a distinctive, concentric, geometric sequence of
CFs which trace the expansion of the virial shock.
\end{abstract}


\section{Introduction}
Recent X-ray observations of galaxy clusters reveal various phenomena
in the gaseous haloes of clusters, such as mergers, cavities, shocks
and cold fronts (CFs).  CFs are thought to be contact discontinuities,
where the density and temperature jump, while the pressure
remains continuous (up to projection/resolution effects).  They are
common in clusters \citep{markevitch07} and vary in morphology (they
are often arcs but some linear or filamentary CFs exist), contrast
(the density jump is scattered around a value of $\sim 2$), quiescence
(some are in highly disturbed merging regions, and some in smooth,
quiet locations), and orientation (some are arcs around the cluster
centres, and some are radial, or spiral in nature).  CFs have been postulated to
originate from cold material stripped during mergers \citep[][ for
  example]{markevitch00} or sloshing of the intergalactic medium
\citep[IGM; ][]{markevitch01,ascasibar06}.
In some cases, a metallicity gradient is observed across the CF,
indicating that it results from stripped gas, or radial motions of gas
\citep[see ][and references therein]{markevitch07}.  Shear is often
found along CFs in relaxed cluster cores, implying nearly sonic bulk flow
beneath (towards the cluster's core) the CF \citep{keshet10}.  Whether
CFs are present at very large radii ($\gtrsim 500~{\rm kpc}$) is
currently unknown because of observational limitations.

Cold fronts are not rare. \citet{markevitch03} find CFs in more than
half of their cool core clusters, suggesting that most, if not all
such clusters harbor CFs \citep{markevitch07}. \citet{ghizzardi10}
survey a sample of 42 clusters using XMM-Newton 
and find that  19 out of 32 nearby ($z<0.075$) clusters host at least one CF. Based on 
observational limitations, and orientations of cold fronts, they
conclude that this is a lower limit. While many CFs are directly
attributed to mergers, out of $23$  clusters in their
sample that seem relaxed, $10$ exhibit
cold fronts. These 10 have systematically lower 
central entropy values. Another useful compilation of clusters for
which cold fronts have been observed can be found in
\citet{owers09}. Out of the 9 high-quality Chandra observations of clusters
with observed cold fronts, 3 of them, RXJ1720.1+2638, MS1455.0+2232
and Abel 2142 have non-disturbed morphologies.

In what follows, we propose that some CFs are produced by merging of
shocks.  When two shocks propagating in the same direction merge, a CF is always
expected to form.  Its parameters can be calculated by solving the
shock conditions and corresponding Riemann problem.  Most of the
scenarios in which shocks are produced (quasars, AGN jets, mergers)
predict that shocks will be created at, or near, a cluster centre, and
expand outwards (albeit not necessarily isotropically).  When a
secondary shock trails a primary shock, it always propagates faster (supersonically with
respect to the subsonic downstream flow of the primary shock), so
collisions between outgoing shocks in a cluster are inevitable if they
are generated within a sufficiently short time interval. We note that
contact discontinuities can also form from head on collisions between
shocks, and various other geometrical configurations. We restrict
ourselves in this paper to merging of two shock propagating in roughly
the same direction.

In \S~\ref{sec:riemann} we solve for the parameters of CFs that are
caused by merging two arbitrary planar shocks. We show results of
shock-tube hydrodynamic simulation that agree with the calculated
analytical prediction. In \S~\ref{sec:spherical}, we describe our spherical
hydrodynamic simulations and realistic cluster profiles that are later
used to investigate
how these
shocks may be produced, and discuss various ways to produce shocks in
cluster settings. We identify two general mechanisms that
create shocks in clusters: energetic explosions near the centre, or
changes and recoils in the potential well of the cluster that
correspond to various merger events.
These mechanisms are simulated in \S~\ref{sec:explosions} and \S~\ref{sec:mergers} 
and also serve to demonstrate the robustness of the shock
induced cold fronts (SICFs) that form.
In \S~\ref{sec:virial} we follow cluster growth and accretion over a
Hubble time by using self-consistent 1D hydrodynamic simulations of gas and
dark-matter. The virial shock that naturally forms serves as a primary
shock, and we demonstrate that SICFs form when secondary shocks
propagating from the centre merge with it. We show that oscillations of this
shock around its steady state position can produce a series of SICFs. 
In \S~\ref{sec:stabil_observe} we discuss aspects regarding the stability and
sustainability of cold fronts in general, and derive some
observational  predictions that could serve to distinguish between
CFs produced via this and other mechanisms. In \S~\ref{sec:summary}
we discuss our findings and conclude. Compact expressions for SICF
parameters can be found in appendix \S~\ref{sec:gamma53}.

\section{Merging of Two Trailing Shocks}
\label{sec:riemann}

In this section we examine two planar shocks propagating in the same
direction through a homogeneous medium and calculate the parameters of
the contact
discontinuity that forms when the 
second shock (secondary) overtakes the leading (primary) one.  This problem is fully
characterized (up to normalization) by the Mach numbers of the primary
and secondary shocks, $M_0$ and $M_1$.  At the instant of collision, a
discontinuity in velocity, density and pressure develops,
corresponding to a Riemann problem \citep[second case
in][\S93]{landau59}.  The discontinuity evolves into a (stronger)
shock propagating in the initial direction and a reflected rarefaction
wave, separated by a contact discontinuity which we shall refer to as
a shock induced CF (SICF).  The density is higher (and the entropy
lower) on the SICF side closer to the origin of the shocks.  In
spherical gravitational systems, this
yields a Rayleigh-Taylor stable configuration if the shocks are
expanding outwards.  Now, we derive the discontinuity parameters.

\subsection{The Discontinuity Contrast}

\begin{figure}
\includegraphics[width=3.5in, trim =190pt 82pt 200pt 30pt,clip=true ]{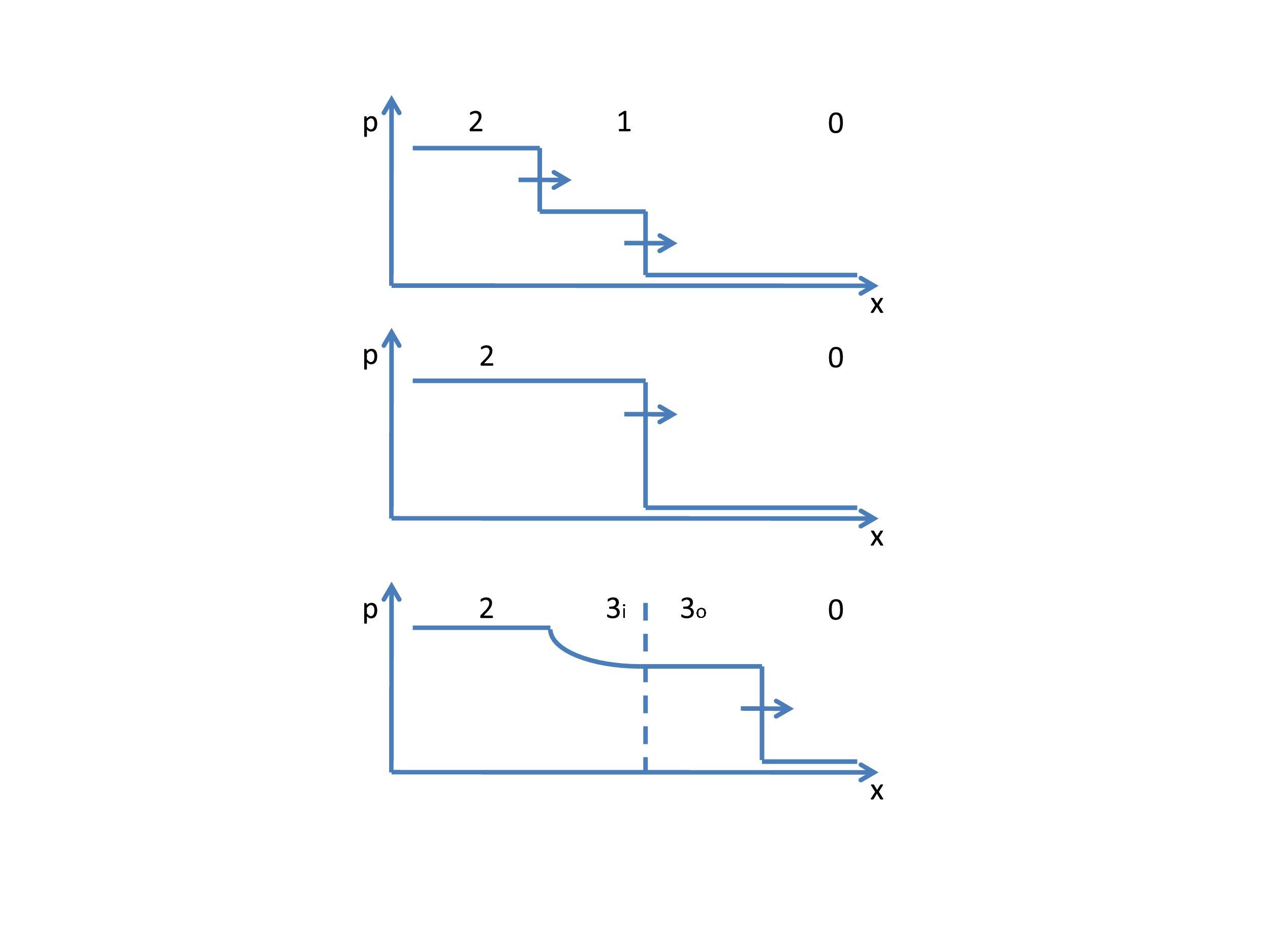}
\caption{Schematics of shock merging. Pressure p is shown vs. an
  arbitrary spatial coordinate x, which is not fixed between the
  panels. {\it Top:} two shocks, propagating according to the arrows,
  induce transitions between zones $0\rightarrow 1$ and between $1\rightarrow 2$. {\it Middle
  panel}, the instance of collisions between the shocks. The transition
  $0\rightarrow 2$ does not correspond to consistent shock jump
  conditions. {\it Bottom:} at the Lagrangian location of the
  collision an isobaric discontinuity between $3_\myout$ and $3_\myin$
  forms, separating between a forward moving shock 
  ($0\rightarrow 3_\myout$) and a reflected rarefaction (smooth transition
  between $2$ and $3_\myin$).\label{fig:scheme}}
\end{figure}

We consider an ideal gas with an adiabatic index $\gamma.$ Before
the shocks collide, denote the unshocked region as zone $0$, the
region between the two shocks as zone $1$, and the doubly shocked
region as zone $2$ (these definitions are demonstrated schematically
in fig. \ref{fig:scheme}).  After the merging, regions $0$ and $2$ remain
intact, but zone $1$ vanishes and is replaced by two regions, zone
$3_\myout$ (the outer region for outgoing
spherical shocks; adjacent to zone $0$) and zone $3_\myin$ (inner; adjacent to $2$),
separated by the SICF.  We refer to the plasma density, pressure,
velocity and speed of sound respectively as $\rho$, $p$, $u$ and $c$.
Velocities of shocks at the boundaries between zones are denoted by
$v_i$, with $i$ the upstream zone.  We rescale all parameters by the
unshocked parameters $\rho_0$ and $p_0$.

The velocity of the leading shock is
\begin{eqnarray}
v_0&=&u_0+M_0~c_0,\label{eq:v0}
\label{eq:v}
\end{eqnarray}
where
\begin{eqnarray}
c_i&=&\left(\frac{\gamma p_i}{\rho_i}\right)^{1/2}
\end{eqnarray}
for each zone $i$.  Without loss of generality, we measure velocities
with respect to zone $0$, implying $u_0=0$.

The state of zone $1$ is related to zone $0$ by the Rankine-Hugoniot conditions,
\begin{eqnarray}
p_1&=&p_0\frac{2\gamma M_0^2-\gamma+1}{\gamma+1}, \label{eq:purho1}\\
u_1&=& u_0 + \frac{p_1-p_0}{\rho_0(v_0-u_0)}, \label{eq:purho2}\\
\rho_1&=&\rho_0\frac{u_0-v_0}{u_1-v_0}. \label{eq:purho3}
\label{eq:purho}
\end{eqnarray}
The equations are ordered such that each equations uses only known
variables from previous equations.
The state of the doubly shocked region (zone $2$) is related to zone
$1$ by reapplying equations \ref{eq:v}-~\ref{eq:purho} with the
subscripts $0,1$ replaced respectively by $1,2$, and $M_0$ replaced by
$M_1$.

Across the contact
discontinuity that forms as the shocks collide (the SICF), the
pressure and velocity are continuous but
the density, temperature and entropy are not; the CF contrast is
defined as $\myr\equiv\rho_{3\myin}/\rho_{3\myout}$.  Regions $0$ and
$3_\myout$ are related by the Rankine-Hugoniot jump conditions across
the newly formed shock,
\begin{eqnarray}
p_3&=&p_0\frac{(\gamma+1)\rho_{3\myout}-(\gamma-1)\rho_0}{(\gamma+1)\rho_0-(\gamma-1)\rho_{3\myout}}, \label{eq:u2u0a}\\
u_3-u_0&=&\left[(p_3-p_0)\left(\frac{1}{\rho_0}-\frac{1}{\rho_{3\myout}}\right)\right]^{1/2} \,. \label{eq:u2u0b}
\end{eqnarray}
The adiabatic rarefaction from pressure $p_2$ down to $p_3=p_{3\myin}=p_{3\myout}$ is determined by
\begin{eqnarray}
u_2-u_3&=&-\frac{2c_2}{\gamma-1}\left[1-\left(\frac{p_3}{p_2}\right)^{(\gamma-1)/2\gamma}\right] . \label{eq:u2u0c}
\end{eqnarray}
The system can be solved by noting that the sum of
eqs. (\ref{eq:u2u0b}) and (\ref{eq:u2u0c}) equals $u_2-u_0$, fixing
$\rho_{3\myout}$ as all other parameters are known from
eqs. (\ref{eq:v0}-\ref{eq:u2u0a}).

Finally, the Mach number of the new shock and the rarefacted density are given by
\begin{eqnarray}
M_f^2 & = & \frac{2\rho_{3\myout}/\rho_0}{(\gamma+1)-(\gamma-1)\rho_{3\myout}/\rho_0} \,, \\
\rho_{3\myin}&=&\rho_2(p_3/p_2)^{1/\gamma} . \label{eq:rho3}
\end{eqnarray}

\begin{figure}
\includegraphics[width=3.5in]{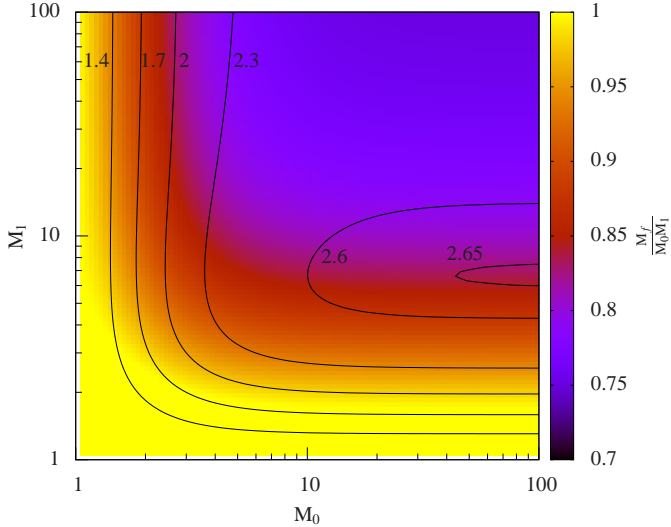}
\caption{Contrast $\myr$ (labeled contours) of a CF generated by a
  collision between two trailing 
  shocks with Mach numbers $M_0$ and $M_1,$ for $\gamma=5/3$ .
  The color map shows the final Mach number $M_f$ normalized by $M_0~M_1.$\label{fig:m0m1}}
\end{figure}
Closed-form expressions for $\myr$ and $M_f$ are presented in \S~\ref{sec:gamma53}.
Figure \ref{fig:m0m1} shows the discontinuity contrast $q$ for various
values of $M_0$ and $M_1,$ for $\gamma=5/3.$  Typically $\myr\sim 2,$
ranging from $1.45$ for $M_0=M_1=2$, for example, to
$\myr_{max}=2.653$. The figure colorscale shows the
dimensionless factor $f\equiv M_f/(M_0 M_1)$, ranging between
$0.75$ and $1$ for $1<\{M_0,M_1\}<100$.

\begin{figure}
\includegraphics[width=3.5in]{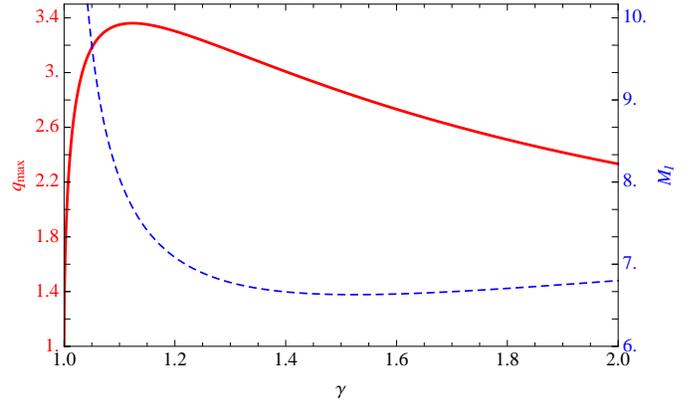}
\caption{ The maximal SICF contrast (solid, left axes) as a function of $\gamma$.
It occurs in a collision between a very strong shock $M_0\to \infty$ and a trailing shock of finite Mach number $M_1$ (dashed, right axes).
For $\gamma= 5/3$, $\myr_{max} = 2.65$ with $M_1 = 6.65$.\label{fig:qmax}
}
\end{figure}
The maximal contrast $\myr_{max}$ depends only on the
adiabatic index, as shown in fig. \ref{fig:qmax}.
It occurs for a very strong primary shock $M_0\to\infty$ and a finite secondary $M_1$;
see \S~\ref{sec:gamma53} for details.
For $\gamma=5/3$ we find $\myr_{max}=2.653$, achieved
for $M_0\gg 1$ and $M_1\simeq 6.65$.

\subsection{Planar Hydrodynamic Example of SICF Formation}
\label{sec:planar}
\begin{figure}
\includegraphics[width=3.6in, trim=35 0 0 65, clip=true]{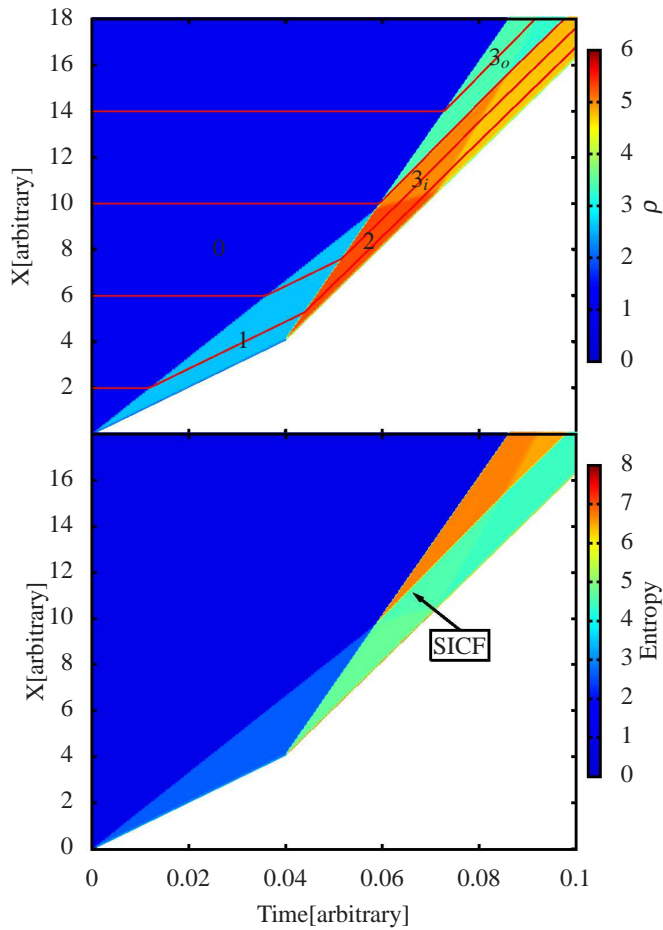}
\caption{Shock tube simulation of a contact discontinuity (SICF)
  formed by merging of two shocks. The piston propagates from $x=0$ in the positive
  direction at a supersonic speed creating the shock between
  $0\rightarrow 1$. At t=.04 the piston's velocity
  abruptly increases causing a second shock to propagate between
  $1\rightarrow 2$. At the position and time where the shocks merge a contact
  discontinuity forms between regions $3_\myout$ and $3_\myin$. {\it Top:}
  density (normalized so that $\rho_0=1$). Red lines mark the
  trajectories of some Lagrangian cell boundaries. The zones here are
  analougous to those in \S~\ref{sec:riemann} and
  fig.~\ref{fig:scheme}. {\it Bottom:} entropy
  (normalized so that $S_0\equiv T_0/\rho_0^{2/3}=1$). Note that there
  is no entropy change between zone $2$ and $3_\myin$ because the
  gas rarefies adiabatically between them. The contrast of the SICF,
  $\myr$, is the density jump between zones $3_\myout$ and
  $3_\myin$. \label{fig:planar}
}
\end{figure}
To demonstrate the formation of an SICF we present in
fig.~\ref{fig:planar} results from a 1D planar hydrodynamic code. The
numerical scheme of the hydrodynamic calculation is Lagrangian, with
the locations and velocities
defined at edges of cells and thermodynamic variables defined at the centres of
cells.  The density $\rho$ and internal energy $e$, are the free thermodynamic
variables, setting the pressure $p$, temperature $T$ and the
entropy $S$ according to an ideal gas equation of state. Throughout
the
paper we consider monoatomic gas, with an adiabatic constant of
$\gamma=5/3$ unless stated otherwise. The time propagation is
calculated by a leap-frog explicit scheme. The timesteps here are
adaptive, and set by a standard Courant Friedrichs Levy
condition. Discontinuities are integrated over by employing
first and second order Von Neumann artificial viscosity.

The simulation presented in fig.~\ref{fig:planar} was run with 500
cells. Without loss of
generality, we can assume $\rho_0=e_0=T_0=1$. The definition of
temperature here implies an arbitrary
heat capacity that does not enter into the equations of motions. We define
the entropy as $S\equiv T/\rho^{\gamma-1}$ (the exponent of the usual
formal definition) so $S_0=1$ as well.
We use external boundary conditions of a piston propagating into the
material from $X=0$ in the positive direction at Mach number
$M_1=2.35$. At time $t=0.04$ the piston's velocity is increased to
create a second shock propagating in the same direction, with
$M_2=1.76$ with respect to the previously shocked gas. The contact
discontinuity that forms has a density ratio\footnote{The SICF
  strength is calculated in the simulation according to
  $\myr=(S_{3\myin}/S_{3\myout})^{-1/\gamma}$ because in the rarefied zone, $3_\myin$ the
  density is not constant, and so harder to measure directly, while
  the entropy is constant throughout the zone.} $\myr\equiv
\rho_{3\myin}/\rho_{3\myout}=1.466$ while the theoretical prediction
according 
to equations \ref{eq:v}-\ref{eq:rho3} is $\myr=1.459$ - in satisfactory
agreement for the resolution used, and for the accuracy needed in
the discussion presented here. The
same technique has also been performed for larger Mach numbers and
yielded results with comparable accuracy.

\section{Numerical Investigation of Spherical Cold Fronts }
\label{sec:spherical}
\subsection{Driving Shocks}
We explore different ways to drive central shocks through clusters of
galaxies. The first, most intuitive method is external energy
injection at the centre (i.e. explosions), which would correspond, for
example, to mechanical energy released in 
AGN bursts. Observations in X-ray and radio indicate a wide variety of
shocks and sound waves \citep{fabian03b,forman07} emitted repeatedly,
perhaps even periodically from a central AGN.
Many clusters also include hot and diffuse bubbles, that are most likely inflated
by AGN jets that are decollimated by the ambient cluster gas
\citep[see review by ][and reference
therein]{mcnamara07}.
If these bubbles are a plausible mechanism to counteract the
overcooling problem, the energies associated with them must be
comparable to the cooling rate of the clusters \citep[$\gtrsim 10^{44}{\rm
  erg~sec}^{-1}$; ][]{edge90,david93,markevitch98} over some fraction
of the Hubble time, indicating that the total energy injection is
$\gtrsim 10^{61}{\rm erg}$.
 The process of inflation of these bubbles sends
shocks through the ICM, but at the same time the extremely hot,
underdense bubbles alter the radial
structure of the ICM in a non trivial manner. In \S~\ref{sec:explosions} we
demonstrate (noting the limitations of the spherically symmetric
analysis in that case) that mergers between shocks driven by such central
explosions can produce SICFs.

Shocks can also be driven gravitationally. Mergers of galaxies and subhaloes
change the structure and depth of the potential well causing sound
waves and shocks to propagate through the haloes. These shocks
typically do not considerably disturb the ICM structure, leaving a
static  gaseous halo after they propagate. We show in
\S~\ref{sec:mergers} that abrupt changes to the gravitational
potential of the halo directly produce shocks. In particular, in the
first passage stage of a merger, two outward propagating shocks are
produced: one when the halo contracts due to the increase of the
gravitational force, and another when the halo abruptly rarefies once
the merging subhalo flies out. The numerical investigation of explosion and merger
driven SICF is performed using a spherical one
dimensional hydrodynamic code within a static potential well
\citep[a version of the code ``Hydra'' documented in ][that is stripped from the
dynamic dark matter evolution and self-gravitating gas]{bd03} that,
along with the initial conditions, is described in
\S~\ref{sec:hydra}.

In \S~\ref{sec:virial} we investigate the interaction of a secondary
shock with the virial shock that is always expected at edges of
clusters. To produce this primary shock, we simulate the
cosmological evolution of a cluster by using a hydrodynamic
scheme that includes a self-gravitating gas that flows within self-gravitating,
dynamic dark matter shells, and starts from spherical cosmological
overdensities at a redshift $z=100$. We show that perturbations can either be
invoked manually, or occur naturally when infalling dark matter has
sufficiently radial orbits so it interacts with the central core
directly. In this context, we also show that low amplitude
perturbations in the core send sound waves that can steepen into a
shock that interacts with the virial shock.

\subsection{The ``Hydra'' Hydrodynamical Code}
\label{sec:hydra}
The cosmological implications of shock interactions in clusters are tested
using 1D, spherical hydrodynamic simulations performed with
``Hydra'' \citep{bd03}.
The code is run here in two modes: cosmological and static DM (dark
matter) modes. In \S~\ref{sec:staticdm} we
use the code in the static DM mode, a simplified
configuration in which the baryons do not self-gravitate, and there
is no evolution of the dark matter. Rather, baryons start
in hydrostatic equilibrium within a fixed cluster-like potential well. In
\S~\ref{sec:virial} we use the code in its cosmological mode,
utilizing the full
capabilities of the code to evolve baryons and dark matter
self-consistently from a cosmological initial perturbation at high
redshift to $z=0$. Next, we
summarize the physical processes implemented in both modes
of ``Hydra'' and then highlight the difference in
boundary conditions and initial conditions corresponding
to each modus operandi.

Similar to the planar code described in
\S~\ref{sec:planar}, the code uses a Lagrangian scheme. The
positions of baryonic spherical shells are defined by their boundaries (with the
innermost boundary at $r=0$), and the thermodynamic properties are
defined at centres of shells. A summary of the fluid equations that
are solved,
and the numerical difference scheme is available in \citet{bd03}.
The code (in its cosmological mode) also evolves thin discrete dark
matter shells, that propagate
through the baryons, and interact with them gravitationally. The full
discrete force equation for the ``i''th baryonic shell is thus:
\begin{equation}
\ddot{r_i}=-4\pi r^2\frac{\Delta P_i}{\Delta
  M_i}-\frac{GM_i}{(r_i+\epsilon)^2}+\frac{j_i^2}{r_i^3},
\label{eq:force}
\end{equation}
with $r_i,M_i,j_i$ the radius of boundary $i$, the mass (baryonic +
DM) enclosed
within that radius, and the specific angular momentum prescribed to
this shell respectively, and $\Delta P_i,\Delta M_i$ the pressure
difference and baryonic mass difference between
the centres of shells $i+1$ and
$i$. $\epsilon$ is the gravitational smoothing length ($50~{\rm pc}$ and
 $500~{\rm pc}$ in the static DM and cosmological modes respectively).

The dark matter and baryonic shells are assigned
angular momentum that acts as an outward centrifugal force designed
to repel shells that are close to the singularity at the centre, in
analogy to the angular momentum of a single star, or of
the gaseous or stellar disk in realistic systems, keeping them from falling into the centre
of the potential well in the absence of pressure support. In
the cosmological mode the shells are initially expanding with a
modified Hubble expansion, and the angular momentum is added as each shell
turns around, corresponding to the physical generation of angular
momentum via tidal torquing around the maximal expansion radius. In
the static DM mode, the angular momentum is present from the beginning
of the simulation, and is taken into account when the hydrostatic
initial profile is calculated. The angular momentum prescription of
the baryons is determined by the requirement that as gas passes
through the virial radius, its angular momentum will be some fraction
\citep[$\lambda=0.05$,][]{bullock01_j} of the value needed to support
it on circular motions. As
expected, the gas only becomes angular momentum supported near the
centre, and when gas cooling is turned on (in the simulations
presented below, the angular momentum of the baryons is never important, and
is presented here only for completeness). The dark matter is
prescribed angular momentum according to the eccentricity of a shell's
trajectory
as it falls in through the virial radius: $j=v_{\rm vir}r_{\rm vir}\sqrt{2(1-\beta)},$
\citep[corresponding to: $v_\theta^2=v_\phi^2=(1-\beta)v_r^2$;][
\S~4.2]{bt87}. $v_{\rm vir},r_{\rm vir}$ are the virial velocity and
radius, and $v_r,v_\theta,v_\phi$ are the radial and two tangential
velocity components of an infalling trajectory. 
Low values
of $\beta$ indicate large angular momentum, and $\beta=1$ corresponds
to purely radial orbits. $\beta$ is constant for all the dark matter shells in the
simulation. Gas cooling can be turned on, by interpolating from a
tabulated version of \citet{sutherland93}, with a prescribed metallicity.

Rather than use a leap-frog integration scheme, which is sufficient for most explicit
hydrodynamic schemes, we use here 4th order Runge-Kutta integration
over the coupled dark matter and baryonic timestep. This allows us to
derive a timestep criterion for the dark matter shells, by comparing
changes in the
velocities and radii of the dark matter and baryons between
the 1st and 4th order propagation and
limit this difference to some fraction of the velocity and radius. The
timestep is determined by the combined Courant Friedrichs Levy
condition and this requirement. If the criterion is not met, the
code reverts to the beginning of the timestep and recalculates the
next step with a reduced timestep.

The numerical integration scheme,
angular momentum and cooling is described in \citet{bd03}. In
addition, we report here on two new ingredients to the calculation:
adaptive changes to the grid, and 1D convective model.
Baryonic shells are now split and merged if they become too thick or
thin respectively. A shell is split if its width is larger
than some fraction (typically $0.2$) of its radius, or if it is larger
than some fraction (typically $4$) of the shell directly below it and
above it, provided that it is larger than a
minimal value (typically $2~{\rm kpc}$), and that the shell is not in
the angular momentum
supported region near the centre of the halo. Shells are merged when the
cumulative width of two adjacent shells is smaller than some
value (typically $0.1~{\rm kpc}$). Shell merging is helpful in cases
where two shells are pressed
together, decreasing the timestep to unreasonable values (reasonable
timesteps for cosmological simulations are above $10^{-7}~{\rm Gyr}$).
An additional component included in ``Hydra'' for the first time is a 1D mixing
length convection model, that acts to transfer energy outwards in regions where
the entropy profile is non-monotonic \citep{spiegel63}. This model
is described in detail in \citet{bd10}, and is used here only in the final
case described in \S~\ref{sec:virial} in its maximal form: bubbles can
rise until they reach the local speed of sound. In this work, the
inclusion of the maximally efficient mixing length theory convection ensures
that the entropy is always monotonically increasing, in an energetically
self-consistent way.

The initial conditions for a cosmological run are derived by requiring an
average mass evolution history for a halo with some final mass \citep{neistein06}. The
overdensity of each shell at the initial time is calculated so that the shell would pass
through its virial ($r_{180}$) radius at the required time \citep[the
procedure is
described in detail in][]{bdn07}. This procedure has an additional
benefit of setting the initial grid in a way that would ensure that
shells will pass through the virial radius at constant time intervals.
In this procedure, and in the
cosmological simulation, a force corresponding to the cosmological
constant has been added to
eq.~\ref{eq:force}, making the simulation completely consistent with
$\Lambda$CDM \citep{bdn07} (the cosmological parameters used
throughout the paper are:
$\Omega_m=0.3,~\Omega_\Lambda=0.7,~h=0.7$). This has little overall
effect, and essentially no
impact once the shells are collapsed, and the overdensities become non-linear.

In the stripped down, static DM mode, there are no dark matter shells,
and the gas does not self-gravitate. Instead, the value $M_i$ of
eq. \ref{eq:force} is interpolated from a predefined lookup table.
In this mode, a rigid external boundary condition is applied, in
contrast to a vacuum boundary condition in the cosmological mode. The
inner boundary condition in both cases is trivially satisfied because
all terms in eq. \ref{eq:force} are individually zero there.

\begin{figure}
\includegraphics[width=3.6in, trim=30 0 10 0, clip=true]{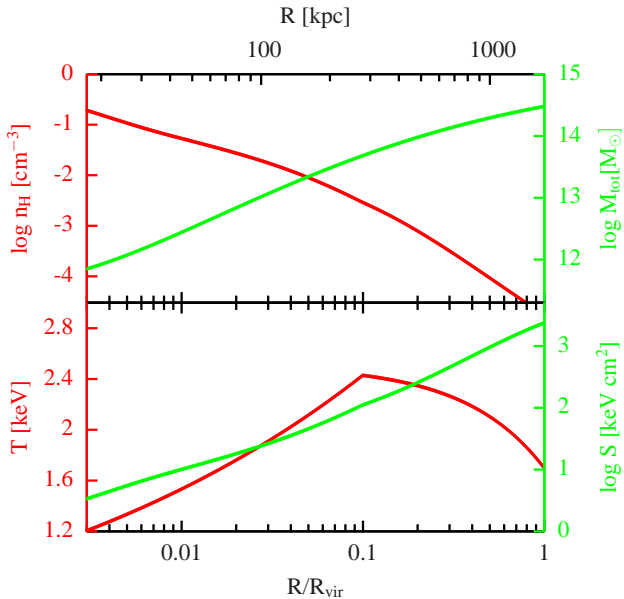}
\caption{Radial profiles of cluster initial conditions. {\it Top:}
  density (red, left axis) and total enclosed mass (green, right axis) of
  the cluster. {\it Bottom:} Temperature (red, left axis) and entropy
  (green, right axis), as a function of radius (upper x axis) and
  virial fraction (lower x axis). The total mass profile, and the temperature
  profile are prescribed, as well as a pivot point in the entropy
  (value of $10~{\rm keV~cm^2}$ at $r=0.01R_{\rm vir}$). The density
  and entropy profiles are calculated so that the gas is in
  hydrostatic equilibrium. The
  temperature profile is set to exhibit a decrease by a factor
  $\approx 1.5$ between $0.1$ to $0.01R_{\rm vir}$.\label{fig:profile}
}
\end{figure}
For the static DM simulations we wish to define initial conditions
which are typical of those in clusters. The total mass profile is
defined by an NFW profile \citep{nfw97}, with
$M_{\rm vir}=3\times 10^{14}M_\odot,$ and concentration set to $c=5.4$ according
to \citet{bullock01_c}. The virial radius is $1730~{\rm kpc}.$ A central galaxy (BCG) is
modeled by a Hernquist profile \citep{hernquist90} superimposed on the
NFW profile, with a sharp cutoff at $10~{\rm kpc}$ and concentration
$a=6~{\rm kpc}$ normalized to $M_{BCG}=3\times 10^{11}M_\odot.$ We define an
inward decreasing temperature profile \citep{leccardi08} that  peaks
at $r_T=0.1~R_{\rm vir}$ at $T(r_T)=T_{\rm
  vir}=2.7\times 10^7K$ and drops inwards
according to $T(r)=T_{\rm vir}(r/r_T)^{0.2}$. Outwards, it is linearly
decreasing from $T(r_T)=T_{\rm vir}$ to $T(R_{\rm vir})=0.7~T_{\rm
  vir}.$ Once the
entropy of a single point is defined, the
density profile can be constructed by the hydrostatic requirement
without further degeneracy.
We set $S(0.01R_{\rm vir})=10~{\rm keV~cm^2}$ as our pivot point \citep{cavagnolo09}. The
final baryonic mass of the resulting profile in the ICM is $2.5\times 10^{13}M_\odot,$
which is $\approx 8\%$ of the total mass. The
density, total mass, temperature and entropy profiles are plotted in
fig.~\ref{fig:profile}. We note that while the constructed initial
profile reasonably
resembles that of a realistic cluster, the mechanism that is studied
here, of SICFs, is not sensitive to these details; any merging between
shocks, in any profile, should produce them.

\section{Static DM simulations}
\label{sec:staticdm}
\subsection{Cluster Explosions}
\label{sec:explosions}
\begin{figure}
\includegraphics[width=3.6in, trim=35 0 0 65, clip=true]{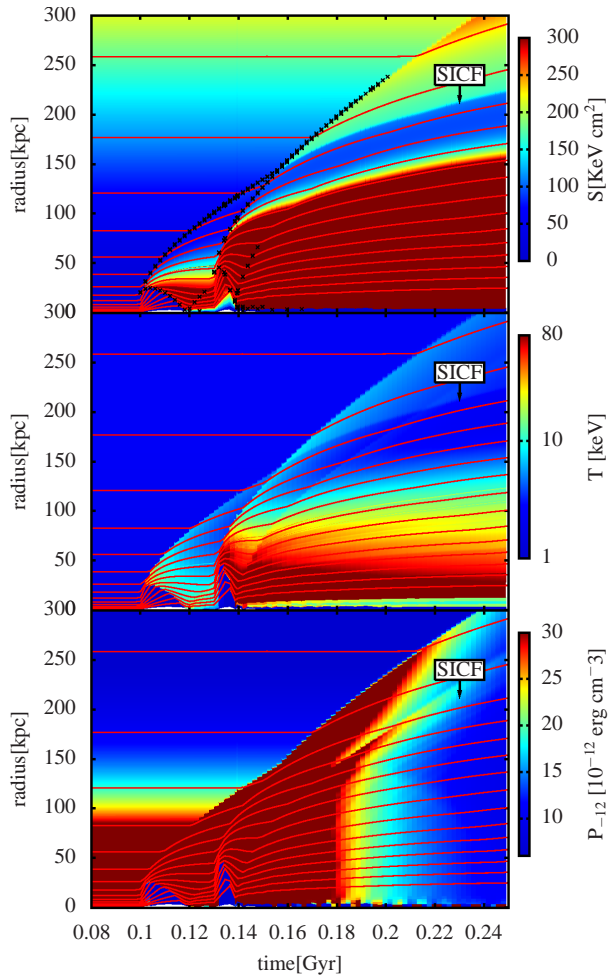}
\caption{Time evolution of an initially hydrostatic halo showing 
  explosion driven shocks from the centre propagating outwards. The
  energy injection parameters are described in the text. {\it Top:}
  entropy color map on the radius-time plane. The red thin lines mark
  the position of every 50th Lagrangian shell boundary. The black
  dots mark 
  shocks, defined as artificial viscosity pressure exceeding
  $10^{-11}erg~cm^{-3}$. These shocks are driven by two explosions, at $t=0.1~{\rm Gyr}$ and
  $0.13~{\rm Gyr}$. The entropy of the shocked gas is saturated
  in this colormap, but the
  entropy jump of the SICF is visible at the Lagrangian location of collision
  between the shocks. {\it middle:} temperature colormap of the same
  simulation. The temperature increase induced by the shocks, and the
  temperature jump at the SICF (by a factor of $1.3$) is visible. {\it Bottom:} Pressure
  colormap. The pressure across the SICF is smooth, as expected for
  cold fronts. \label{fig:explode}
}
\end{figure}
The first mechanism we investigate to produce shocks is that of cosmic
explosions. Many processes can cause abrupt energy injection at
centres of clusters, among which are AGNs or starbursts at the BCG and
galaxy-galaxy mergers. The discussion in this section is confined to
injection to the baryonic component itself (changes to the
gravitational potential well will be discussed next). As shocks sweep
through matter,
they constantly loose energy to heating the gas, and when the shocks
expand spherically, they also weaken due to the geometrical increase
in the surface of the shock \footnote{\citet{waxman93,kushnir05}
  showed that for power law density profiles and strong shocks, as
  long as the mass diverges with radius, or, $\rho_{\rm gas}\propto
  r^{-w},$ with $w\leq 3$, the shock will weaken}. We 
seek here to create shocks that will survive to some considerable 
fraction of the halo
radius, requiring that the amount of energy that is injected will be
comparable to the total thermal energy of the gas in the cluster. This
energy scale is similar to that required to solve the overcooling
problem, indicating that any solution to the cluster overcooling
problem via strong shocks will produce shocks that will expand to a
significant fraction of the halo radius.

Fig.~\ref{fig:explode} describes the time dependent
evolution of an initially hydrostatic cluster atmosphere, as two
explosions are driven through it. The simulation has $1,000$
logarithmically spaced shells, between $0.0005~R_{\rm vir}$ and
$R_{\rm  vir}.$ Here, and in the following examples, the cooling is
turned off, which is justified in the absence of any feedback
mechanism in the simulation that would counteract the
overcooling. The explosions are driven by setting a homologic
($v=v_{\rm exp}r/r_{\rm exp}$) velocity profile with
highly a supersonic velocity ($v_{\rm exp}$), sharply cutting off beyond some
radius $r_{\rm exp}$. For
the first explosion, these parameters are $v_{\rm exp}=10,000~{\rm
  km~sec}^{-1}$ and $r_{exp}=20~{\rm kpc},$ injecting
$1.7\times 10^{61}~{\rm erg}$ into the core of the cluster. The second
  explosion, $3\times 10^7~{\rm yr}$ later, has a velocity of $v_{\rm exp}=20,000~{\rm km~sec}^{-1}$ at $r_{exp}=40~{\rm kpc},$ and
injects $4.7\times 10^{61}~{\rm erg}.$ These large values are not a-typical for
the energy estimated in radio bubbles \citep{birzan04,mcnamara07}. While no buoyancy is
possible within a spherical symmetric framework, the explosions do
create extremely hot and dilute regions, and drive strong shocks.
The two shocks can be traced in fig.~\ref{fig:explode} by the breaks in the plotted
Lagrangian lines (velocity jump), by
the temperature jump, and by the regions in the simulation with large
artificial viscosity (black dots). At the location of merging
between the shocks (at $t=0.15~{\rm Gyr}$ and radius $\approx 150~{\rm
  kpc}$), an SICF (contact discontinuity) forms (visible in
the entropy colormap) that propagates with the matter. We have
followed the simulation for $1~{\rm Gyr}$ and, as expected in the
absence of any diffusive mechanisms, the SICF retains its strength.

The energies and time interval between the shocks have been chosen arbitrarily to
create a shock collision at $\approx 150~{\rm kpc}.$ A series of
sharp, frequent and strong central shocks is expected in some
self-consistent radiative and
mechanical AGN feedback models \citep{ciotti07,ciotti09}, and is
indicated in some observations \citep[for example, ][for Perseus and
M87 respectively]{fabian03b,forman05}. Non-spherical,
frequent collisions may occur between the shocks produced by
the two bubbles of a radio active AGN, at a position roughly above one
of the bubbles \citep[the generation of two shocks corresponding to observed
bubbles, as well as
radial soundwaves that steepen into a shock is discussed
in][]{fabian03b}. The shock from the near-by bubble (primary) arrives to that position
 first, and
the shock from the opposite bubble (secondary) arrives later. Given our 1D
framework, we do not attempt to
map the parameter
space of possible explosions. Since shocks weaken as they propagate
outwards, weaker shocks which occur within a shorter time interval
will produce smaller radii SICF with the same strength as stronger
shocks driven within a larger time interval.

\subsection{Gravitational Perturbations: Merger Induced Shocks}
\label{sec:mergers}
If the core of an otherwise hydrostatic cluster is gravitationally
perturbed, the reaction of the gas to the perturbation often results
in shocks. While explosions, discussed in \S~\ref{sec:explosions},
typically considerably heat the shocked gas
near the centre, gravitational perturbations make a more subtle  change to
the gas, while forming strong outward expanding shocks. The core is perturbed
when gas or dark matter
is accreted to the centre of the cluster, either smoothly (creating
weak perturbations) or by mergers, (causing a more violent and abrupt
shock). In this section we describe the shocks that occur as a
result of a large change to the core potential depth of a cluster due
to mergers. In \S~\ref{sec:virial} we show that weak oscillations in
the core send sound waves through the gas, that steepen into shocks.

\begin{figure}
\includegraphics[width=3.6in, trim=35 0 0 65, clip=true]{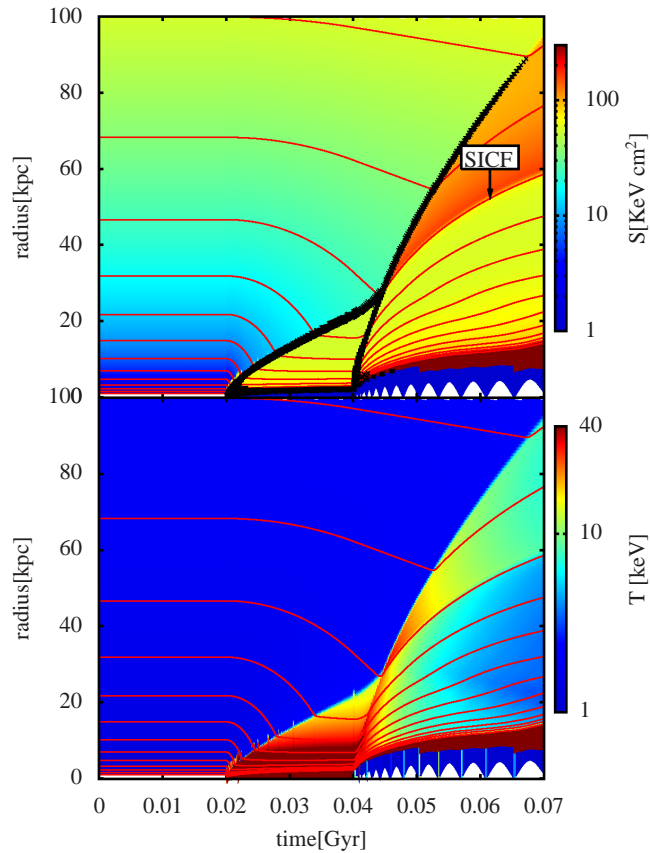}
\caption{Same as fig.~\ref{fig:explode}, but now the
  potential well is modified by artificially adding a central object
  of mass  $3\times 10^{13}M_\odot$, which is then
  removed $2\times 10^7{\rm yr}$ later (corresponding to a first passage of
  a 1:10 ratio subhalo). Note that the scales are different from fig.~\ref{fig:explode}. \label{fig:flyby}
}
\end{figure}
Fig.~\ref{fig:flyby} shows a merger event of a cluster with the same initial conditions
described in \S~\ref{sec:hydra} and fig.~\ref{fig:profile} with a
dense substructure with total mass of 
$3\times 10^{13}M_\odot$ (a $10\!\!:\!\!1$
merger ratio). The
simulation describes, in a simplified form, how the
main halo will react to the first passage of the substructure near its
core. During its first pericenter passage, the subcluster will first
cause the core to contract, and then, as it flies out, allow the core
to relax again. Here, we model this in a simplistic way by adding, and
then removing (after $2\times 10^7~{\rm yr}$), a massive object at the centre of the simulated
cluster. The addition of the mass causes all the cluster's atmosphere to contract,
affecting the inner region more than the outer. A coherent
flow inwards begins, that stops as an outward expanding shock passes
through the cluster, readjusting it to a new hydrostatic equilibrium
(note in fig.~\ref{fig:flyby} that after the passage of the primary shock the post-shock gas
is almost at rest). Once the central object is removed, the halo jolts out from the
inside out (again, because the addition of the central force affects
the inner core the most), causing another shock to
occur. The energy for these two shocks is taken from
the orbital energy of the merging subcluster, and would cause it to
spiral in. In the figure both shocks are clearly visible in the
temperature maps, and through the black dots corresponding to strong
artificial viscosity. At the instant of merging between the shocks (at $t=0.044~{\rm Gyr}$ and radius $\approx 25~{\rm
  kpc}$),
an SICF forms which is seen in the entropy and temperature
colormaps. The time interval for the merger interaction is chosen manually, but
does not seem improbable noting that an object approaching the centre
at $1000{~\rm km~sec}^{-1}$, roughly the virial velocity, passes
$20~{\rm kpc}$ during that period. Shocks associated with sequential
core-passages of an inspiraling subhalo could also induce large scale
SICFs. While shocks produced by merger
events have been seen in simulations \citep{mccarthy07},
the formation of SICF in these simulations hasn't
been tested yet, and is left for future work.

\section{Virial Shocks and SICFs}
\label{sec:virial}
At the edges of galaxy clusters there are virial shocks with typical
Mach numbers $\sim 30-100,$ heating the gas from $\sim 10^4~{\rm K}$
to $\sim 10^7~{\rm K}$ by converting kinetic energy into thermal
energy. Virial shocks are probably observed by Suzaku \citep{george09,hoshino10}.
The rate of expansion of a virial shock is set on average by the
mass flux and velocity of the infalling material
\citep[see for example][]{bertschinger85}. Secondary shocks that originate from
the centre of clusters, due, for example, to the mechanisms discussed in
\S~\ref{sec:spherical}
will collide and merge with the virial shock, creating SICFs at the
locations of the virial shock at the corresponding times. Since
shock mergers with virial shocks require only one additional shock to form
(rather than two in the general cases described in
\S~\ref{sec:staticdm}), and no
particular timing is required, such SICFs are perhaps even more
probable. When the Mach number of the primary (virial) shock is $M_0\gtrsim 3$,
which is always the case,
the strength of the SICF becomes almost independent of the secondary
shock, allowing for a rather narrow range of strengths around
$\myr\approx 2$ as long as the secondary shock is sufficiently strong (fig.~\ref{fig:m0m1}).

We simulate the merging of a virial shock with a secondary by evolving self consistently the dark matter and
baryons from an initial perturbation in the cosmological mode described in \S~\ref{sec:hydra}.
These initial conditions provide a reasonable laboratory for gas dynamics
in the ICM, producing correct temperatures and
densities, and tracking the
virialization process of the gas in a self consistent way. In order to
capture the production of a secondary shock and its interaction with
the virial shock in 1D simulation, we examine various artificial
schemes which mimic the full 3D dynamics, as discussed below. A more
detailed simulation of cluster evolution would require 3D simulations
that include AGN feedback and merger and smooth accretion
for baryons and dark matter.
The cosmological simulations here have been performed with $2,000$
baryonic shells, and $10,000$ dark matter shells, sufficient according
to our convergence tests.

For the average mass accretion histories used here
(\S~\ref{sec:hydra}), the
accretion rate grows with halo mass, even after taking into account
the late time decline in accretion rates \citep{dekel09}, so
shells that fall later are more massive. When late
 massive dark matter shells fall into the inner core, their
mass becomes comparable to that of the gas enclosed within that shell, causing
it to vibrate stochastically. These core vibrations drive sound
waves that are emitted from the centre
and steepen to create shocks (the energy for these
shock is taken from the orbits of the dark matter shells). We can
control the level of stochastic noise by increasing or decreasing the
angular
momentum prescribed to the dark matter. In the first two ``synthetic''
examples shown here, we use $\beta=0.7$ (see \S~\ref{sec:hydra}) to reduce noise in the
evolution, and then manually add perturbations at time $-7~{\rm Gyr}$
($z=0.84$). In the last example, we decrease the dark matter angular
momentum ($\beta=0.991$), and use the numerical stochastic noise
as a proxy for the perturbations that halo cores undergo during their
formation and accretion. We demonstrate that this noise naturally give rise
to shocks that interact with the virial shock. This interaction seems
to be quasi periodic, and we suggest a mechanism for driving this behavior. In
this final example, the entropy profile is sometimes non-monotonic,
implying that convection might be important. We
 test the effect of convection using our 1D mixing length theory and
show that it does not significantly reduce the strength of the SICFs
that are produced, and does not change the overall evolution significantly.

\subsection{Manual Initiation of Secondary Shocks}
\label{sec:manual_shocks}
\begin{figure*}
\includegraphics[width=7.in, trim=40 0 50 0, clip=true]{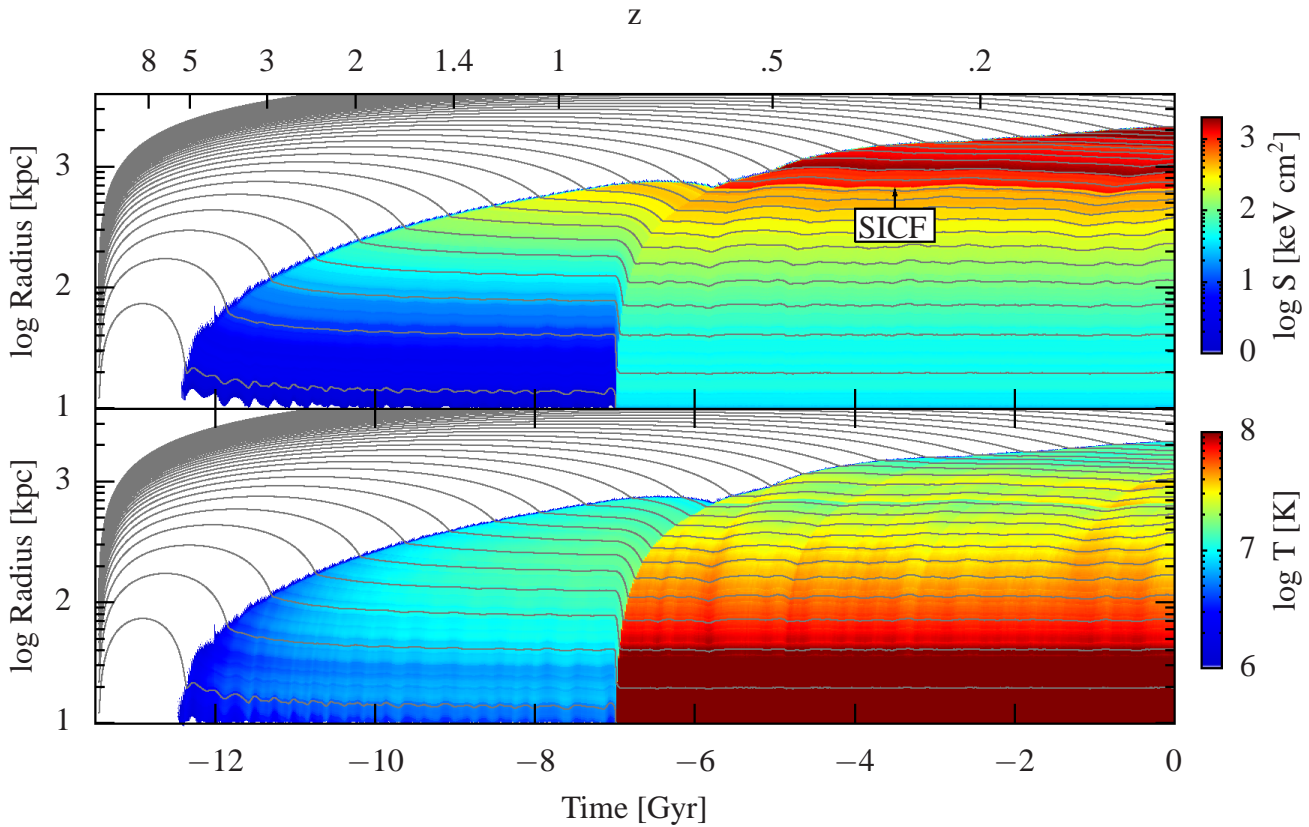}
\caption{Cosmological mode simulation of the evolution of a
  $M(z=0)=3\times 10^{14}M_\odot$ cluster. The gray lines mark the evolution of
  every 50th Lagrangian shell,
  and the colormaps are entropy ({\it top}) and temperature ({\it
    bottom}). The initial Hubble expansion, turnaround and the virial
  shock are visible.
Initially, at $z=100$, shells expand due to the
  near-Hubble flow. At time $-7~{\rm Gyr}$ a central object with mass
  $3\times 10^{13}M_\odot$ is permanently added, causing a shock to
  propagate outwards. The collision between this shock and the virial
  shock creates an SICF visible in the entropy colormap which
  persists to $z=0$.\label{fig:cosmo_madd}
}
\end{figure*}
\begin{figure*}
\includegraphics[width=7.in, trim=40 0 50 0, clip=true]{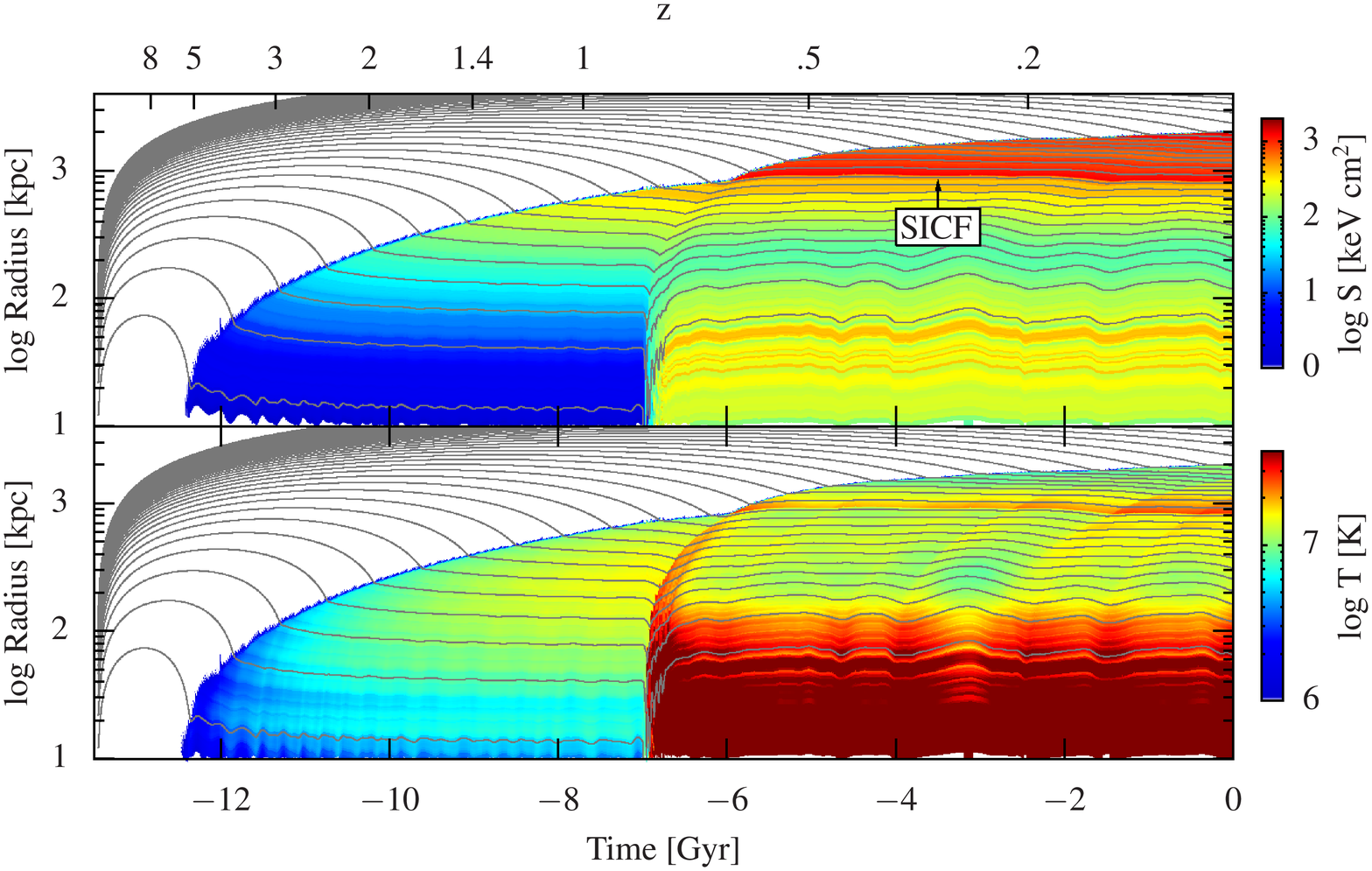}
\caption{A cosmological mode simulation with the same initial
  conditions as in fig.~\ref{fig:cosmo_madd}, but with the core
  perturbed briefly between $-7$ and $-6.75~{\rm Gyr}$ (see
  \S~\ref{sec:manual_shocks} for details). The perturbation
  is a sinusoid in time, with an amplitude of $3\times 10^{13}M_\odot$, and
  a period of $5\times 10^7~{\rm Gyr},$ following by a constant left-over
  mass of  $5\times 10^{12}M_\odot.$ Weak shocks and sound waves are sent
  outwards, steepening into a single shock that collides with the
  virial shock, producing an SICF.\label{fig:cosmo_msin}
}
\end{figure*}
Fig.~\ref{fig:cosmo_madd} describes a typical evolution of a cluster
that ends up with mass
$3\times 10^{14}M_\odot$ at $z=0$, and subsequent interaction of the
virial shock with a
secondary merger induced shock. The merger here is $10\!\!:\!\!1$, with a
central mass of $3\times 10^{13}M_\odot$ instantaneously added at  $-7~{\rm
  Gyr}$ (at that time the cluster's virial mass
is $\approx 10^{14}M_\odot$). Unlike the procedure described in
\S~\ref{sec:mergers}, we do 
not remove the merged mass, creating only one shock. The Hubble
expansion, turnaround and virial shocks are seen,
as well as the SICF formation at time $\approx -5.5~{\rm Gyr}$ and at
radius $\approx 700~{\rm kpc}$. After the deepening of the gravitational
potential well, the secondary shock mediates the re-establishment of
hydrostatic equilibrium. Thus, the post-shock gas, and the SICF, are roughly at
rest in the Eulerian laboratory frame of reference.
The post-shock temperatures, particularly below $100~{\rm kpc}$ are
unrealistically high, corresponding, perhaps, to disturbed morphologies seen in
cluster merger events \citep[the bullet cluster; ][]{tucker98}, although no entropy inversion occurs there. This
is remedied somewhat when the shock is initiated by a series of weaker
shocks and sound waves that steepen into a strong shock further from
the cluster's centre (fig.~\ref{fig:cosmo_msin}), and is completely
alleviated when the shocks are due to persistent stochastic noise near
the core (\S~\ref{sec:natural_shocks}), generating profiles which are
also consistent with quiescent cool core clusters.

We produce a weaker, more local perturbation by engineering a
perturber that sends weak shocks and sound waves that steepen into a
shock.
Fig.~\ref{fig:cosmo_msin} shows an evolution of the same cluster as in
fig.~\ref{fig:cosmo_madd}, but with a central perturber who's mass is
sinusoidally changing between $-7<t<-6.75~{\rm Gyr}$ with an amplitude
of  $3\times 10^{13}M_\odot$  and a period of of $5\times 10^7~{\rm
  Gyr}.$ After that, a constant mass of $5\times 10^{12}M_\odot$ is
left. The steepening of the shocks is visible, and the temperature
increase of the gas by the secondary is much smaller, because,
contrary to the permanent  mass addition in fig.~\ref{fig:cosmo_madd},
no net contraction is caused. This example serves to show how shocks
can form from ``noise'' generated near the core, but still heat the
centre from the inside out, in a way that would correspond to a merger
event but not to a quiescent cool core cluster. The entropy
profile here becomes non-monotonic near the centre.

\subsection{Self Consistent Initiation of Secondary Shocks}
\label{sec:natural_shocks}
\begin{figure*}
\includegraphics[width=7.in, trim=40 0 50 0, clip=true]{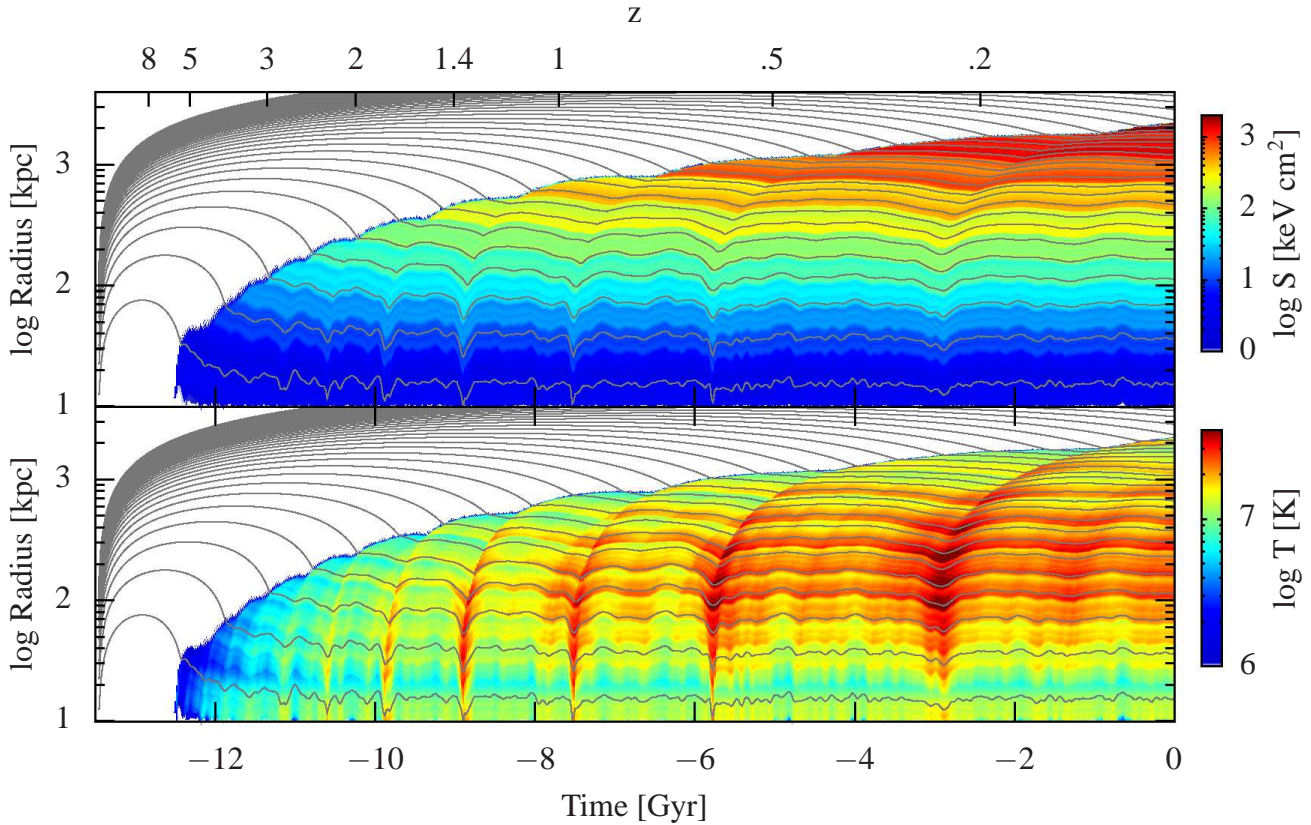}
\caption{A cosmological mode adiabatic simulation of cluster
  evolution with initial conditions as described in the
  \S~\ref{sec:natural_shocks}. The 
  angular momentum of the dark matter is sufficiently small to allow
  massive, late accretion of dark matter shells to propagate and reach
  the centre, increasing the amount of stochastic ``noise'' with
  respect to the levels seen in figures \ref{fig:cosmo_madd} and
  \ref{fig:cosmo_msin}. A series of secondary shocks merge with the
  virial shock, creating a series of SICFs.\label{fig:cosmo_ad}
}
\end{figure*}

\begin{figure}
\includegraphics[width=3.3in]{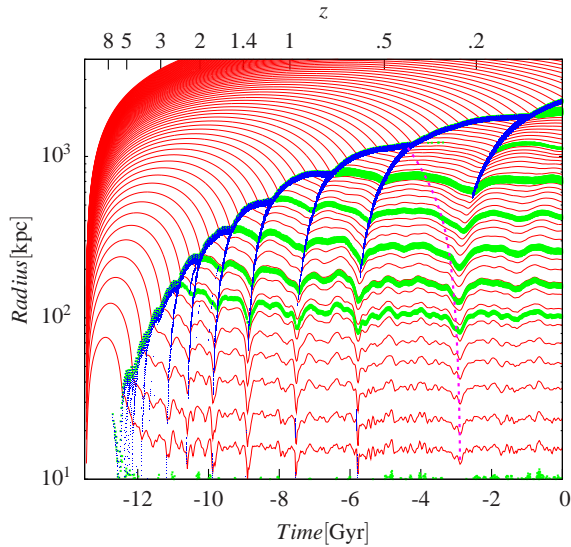}
\caption{Evolution of a galaxy cluster from cosmological initial
  perturbations. The final mass of the cluster at $z=0$ is $3\times
  10^{14}M_\odot.$ The radius and time of Lagrangian shells (every
  $25^{th}$ shell) are  plotted in red thin lines. Shocks are traced by large
  Lagrangian derivatives of the entropy ($d{\rm ln}S/dt>.1~{\rm Gyr}^{-1}$,
  blue dots), and CFs are traced by their large entropy gradients
  ($\partial{\rm ln} S/\partial {\rm ln}r>0.5$, green dots).
  Rarefaction waves can be seen as small motions of the Lagrangian
  shells (illustrated by the dashed magenta curve, manually added
  based on a time series analysis). Their trajectory is approximately
  the reflection of the preceding outgoing compression, time inverted
  about the last secondary-virial shock collision.
  \label{fig:time}}
\end{figure}

\begin{figure}
\includegraphics[width=3.5in, trim =40pt 0pt 50pt 0pt,clip=true ]{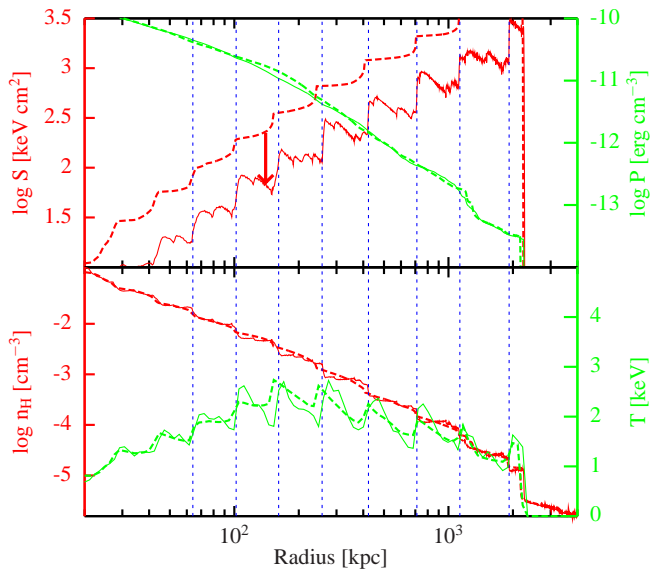}
\caption{Thermodynamic profile of the simulated cluster shown in
  figures \ref{fig:cosmo_ad}-\ref{fig:time} at $z=0$ (solid lines) and of the same
  simulation with maximal convection (dashed lines). {\it Top:}
  entropy (red, left axis) and pressure (green,right
  axis). The entropy of the convective simulation has been raised by
  $0.5~{\rm dec}$ to distinguish between the models. {\it Bottom:} density (red, left axis) and temperature
  (green, right axis). The positions of CFs in the adiabatic, non-convective
  simulations are marked with blue dashed
  vertical lines.\label{fig:prof}
  }
\end{figure}

We allow numerical noise to perturb our hydrodynamic simulation by
reducing the angular momentum of dark matter so that late infalling,
massive dark matter shells penetrate the cluster core. The resulting
time evolution and final profile are shown in figures
~\ref{fig:cosmo_ad}~-~\ref{fig:prof}. As soon as the cluster progenitor
forms, at high redshift, sound waves and weak shocks begin propagating through the
cluster, as the gas constantly adjusts to the varying potential well
near the centre. The weak shocks steepen and accumulate to create
strong shocks that propagate outwards and merge with the virial shock,
creating an SICF on each merger instance. From fig.~\ref{fig:cosmo_ad}
it is evident that the shocks are weak, and that they do not alter the
inner profile considerably.

The shocks that are invoked by the stochastic noise seem to be
periodic, with a period corresponding to about twice the sound (or
shock) crossing time through the cluster. Lacking a formal analysis
for this oscillatory mode, we suggest, based on these simulations, the
following explanation. The location of the virial shock is regulated
such that the
pressure below (towards the cluster's centre) will support the shock
and the ram pressure of the infalling gas
\citep{bertschinger85}. While this process would
imply a smooth expansion of the shock, the virial shock can
oscillate around its steady state trajectory in
response to perturbations in the halo. When the post-shock gas is
over-pressurized, the virial shock adjusts itself by expanding
faster, relaxing this overpressure. If the relaxation overshoots, the
pressure would drop below its steady state value, causing the shock to
decelerate. The halo thus oscillates between these two 
phases depending on the virial 
shock's velocity: in one it is overly slow,
producing pressurization of the
post-shock gas, and in the other it is overly fast, causing
rarefaction. These modes are interlaced. A
rarefaction cycle begins with the secondary hitting the virial
shock sending
rarefaction waves inwards that are reflected at the centre and
propagate outwards, ultimately reaching the virial shock
making it
decelerate. A compression, or secondary shock cycle
collects compression waves that are transmitted inwards when the
virial shock is
too slow, and are reflected through the centre, ultimately steepening
into what would become the secondary shock. Circumstantial evidence
for this mode can be seen in fig.~\ref{fig:time} where the
shocks are automatically traced, and the trajectory of one rarefaction
wave reflected from a secondary-virial shock merger is highlighted by
hand (dashed curve, based on a high resolution time sequence). Note
that this periodicity is also seen (albeit much 
more weakly) in the two
manual excitations of a secondary shock described in figures
\ref{fig:cosmo_madd} and \ref{fig:cosmo_msin}. The periodic SICFs that
are produced can be seen in the entropy profiles in the upper panel of
fig.~\ref{fig:cosmo_ad}, in the lines marking the gradient of the
entropy in fig.~\ref{fig:time}, and in the final profile of this
simulation shown in fig.~\ref{fig:prof}.
The shocks and rarefactions propagating though the halo, and the
alternating speed of the virial shock\footnote{A movie of the evolution of radial
profiles in time is available at
\url{http://www.cfa.harvard.edu/~ybirnboi/SICF/sicf.html} } are observed in the 3D galactic halo simulations of \citet{kh09}\footnote{Du{\v s}an Kere{\v s},
(private communication). Propagation of merger induced shocks and the
subsequent bounce of the virial shock as these secondaries merge with
it are seen in
\url{http://www.cfa.harvard.edu/~dkeres/movies/B1hr_10n128_large_gas.mp4}. There,
left panel is gas density, right panel temperature, and a particularly
notable shock forms at $z\approx 1.7$.}.
Note that this qualitative argument is similar
  to the one described in \S~\ref{sec:riemann}. The overpressure
there is caused by a secondary shock that perturbs the equilibrium
solution of a single shock: the primary accelerates, and a rarefaction is
sent back, relaxing the gas.

 The calculations presented in Figures
\ref{fig:cosmo_ad}-\ref{fig:prof} are adiabatic.  When cooling is
turned on, and
in the absence of feedback, this simulation suffers from overcooling,
creating an overmassive BCG of $2\times 10^{12}M_\odot$ and BCG
accretion rates (corresponding to star
formation rates) of $\gtrsim 100~M_\odot{\rm yr}^{-1}$, and the luminosity
exceeds the $L_x-T$ relation \citep{edge90,david93,markevitch98}. The excitation of the basic mode, and
periodic CFs, are observed in all these simulations. Since the virial shock's
strength is non-monotonic, it is not surprising that between SICF
formation, the post-shock entropy is non-monotonic. This introduces a
problem in 1D simulations where convection due to entropy inversion
cannot naturally occur. We test this by rerunning the adiabatic
simulation with 1D convection according to mixing length theory
\citep{spiegel63}, with the mixing length coefficient set such that
the convection is limited only by the requirement that bubbles never
exceed the sonic speed. This acts to redistribute the energy and
entropy between SICFs, but does not change the overall dynamics. The
final profile in this case is plotted in the dashed lines of
fig.~\ref{fig:prof}. The strength of the SICF in this case is reduced,
but it is still clearly visible, with typical values of $\myr\approx
1.5$. It is highly improbable that convection operates in its maximal physical
efficiency, particularly in the presence of ICM magnetic
fields. \citet{parrish08b,parrish09} show that in the presence of weak
magnetic fields the effective convection is smaller by many orders of
magnitude than the heat fluxes carried by conduction, implying that it
is highly suppressed from its mixing length theoretical value (Ian
Parrish, private communication).

We note a qualitative similarity between reverberations of the virial accretion shocks
discussed here, and another prominent case of accretion shocks in
astrophysics - of type II core collapse supernovae. \citet{burrows07} find that
stalled accretion shocks around type II cores are unstable to 2D
(g-mode) reverberation that,
after enough cycles, accumulate
sufficient energy and amplitude to cause
an outbreak of the stalled shock. While the
standing accretion shock instability (SASI) is predominantly 2D, and
acts on different scales than discussed here, it does draw its
energy from the accreted gas, and is perturbed by sound waves
that are emitted from the vibrating core, that accumulate to a single
unstable mode.

\section{Stability and Observability of Shock Induced Cold Fronts}
\label{sec:stabil_observe}
\subsection{Stability}
\label{sec:stability}
The stability of CFs limits the time duration over which they are
detectable, and so is important when comparing CF formation models
with observations.  Various processes can cause a gradual breakup or
smearing of the CF.  They act in other CF formation models as well.

Thermal conduction and diffusion of particles across the CF smears the
discontinuity on a timescale that is set by the thermal velocity and
the mean free path of the protons. The Spitzer m.f.p. $\lambda$ in an
unmagnetized plasma with typical cluster densities is a few kpc, and
depends on the thermal conditions on both sides of the discontinuity
\citep{markevitch07}. Taking $\lambda \sim 10~{\rm kpc}$ and a thermal
velocity $\bar{v}\sim 1000~{\rm km \, sec}^{-1},$ and assuming that a
CF is visible if it is sharper than $L_{\rm obs}\sim 10~{\rm kpc},$ we
get a characteristic timescale for CF dissipation,
\begin{equation}
t\sim \frac{L_{\rm obs}^2}{D}\sim\frac{3L_{\rm
    obs}^2}{\bar{v}\lambda}\sim 10^7{\rm yr}.
\end{equation}
This result indicates that in order for shock induced CFs to be
observable, either (i) they are formed frequently \citep[e.g., by a
series of AGN bursts; see][]{ciotti07,ciotti09}; or (ii) magnetic
fields reduce the m.f.p. considerably, as some evidence suggests
\citep[see the discussion in][]{lazarian06}.

Heat flux driven buoyancy instability \citep[HBI; ][]{parrish08,quataert08,parrish09} tends to
preferentially align magnetic fields perpendicular to the heat flow in
regions where the temperature decreases in the direction of gravity.
This effect has been argued to reduce radial diffusion within the inward cooling
regions in the cores of cool core clusters.
As suggested in \citet{parrish08}, the temperature gradient across a
CF is opposite to the direction of
gravity, and HBI instability is expected to quickly form there (over $8~{\rm
  Myr}$ in their example), aligning the magnetic fields parallel to
the discontinuity and
diminishing radial diffusion. This possibility needs to be addressed
further by detailed numerical magneto-hydrodynamic simulations.
Also, the timescale for achieving local thermal equilibrium between
the electrons and ions below the virial shocks are long. It is unclear
what the proper diffusion 
coefficients are in that case, and a particle transport analysis needs
to be applied. These important issues are left for future work.

\begin{figure}
\includegraphics[width=3.3in]{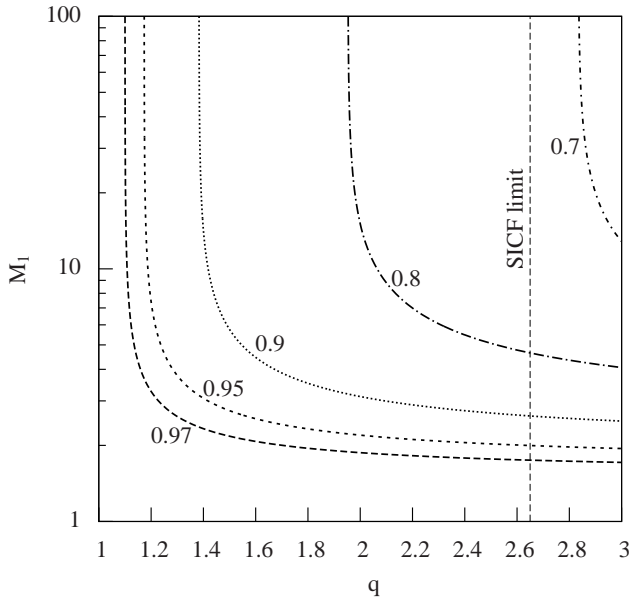}
\caption{Fractional decline (contours) in initial CF contrast $\myr$
  induced by a collision with a shock of Mach number $M_1$ arriving from the dense CF side, for $\gamma=5/3.$
  The vertical dashed line corresponds to the maximal SICF contrast, $\myr_{max}$.
  \label{fig:mr}}
\end{figure}
CFs  could also be degraded by subsequent shocks that sweep outwards across
the CF. This is true regardless of their formation mechanisms.
However, the CF contrast is only slightly diminished by this
effect. This is shown analytically by recalculating the Riemann
problem described in \S~\ref{sec:riemann} but starting from a general
contact discontinuity with strength 
$\myr$ instead of a primary shock. For $\myr=2$ this decrement is between $4\%$ for $M_1<2$, to
$20\%$ for $M_1\gtrsim 10$, as
shown in fig.~\ref{fig:mr}. Such passing shocks seed
Richtmyer-Meshkov instabilities which could cause the CF to break
down.  Although less efficient than Rayleigh-Taylor instabilities,
these instabilities operate regardless of the alignment with gravity.
The outcome depends on $M_1$ and on any initial small perturbations in
the CF surface.  However, the instability is suppressed if the CF
becomes sufficiently smoothed.

It is worth noting that KHI, which could break down CFs formed through
ram pressure stripping and sloshing, does not play an important role
in SICFs because there is little or no shear velocity across them.
However, the stabilizing alignment of magnetic fields caused by
such shear \citep{markevitch07,keshet10} is also not expected here.

\subsection{Observability}
\label{sec:observe}
\citet{owers09} present high quality Chandra observations of
$3$ relaxed CFs. They all seem concentric with respect to the cluster
centre, and spherical in appearance. They interpret these as
evidence for sloshing, and find spiral characteristics in $2$ of
them. In the cold front sample of \citet{ghizzardi10}, $10$ of the
relaxed morphology clusters also host cold fronts. We argue here that
some CFs of this type could originate from 
shock mergers.  In the SICF model, CFs are spherical, unlike the
truncated CFs formed by ram pressure stripping or sloshing, so no
special projection orientation is required.  On the other hand, SICFs
do not involve metallicity discontinuities, observed in some of these
CFs.  The observed contrasts of these three CFs in \citet{owers09} are quite uniform, with best
fit values in the range $2.0-2.1$ (with $\sim 20\%$ uncertainty) in
all three, as expected for SICFs\footnote{However, shocks sweeping
  over CFs could diminish their contrast towards $\myr\sim 2$,
regardless of CF origin.  A larger, more complete sample of CFs is
needed to determine the origin/s of the relaxed population.}.

The following characteristics are specific to SICFs, and could thus be
used to differentiate between SICFs and CFs formed by other
mechanisms. These are also predictions for SICFs that are expected to
form at radii that are not observable today.

{\bf Morphology.} SICFs are quasi-spherical around the source of the
shocks. In contrast, for example, a galactic merger defines an orbit plane, and the
corresponding perturbation (stripped material or sloshing and centre displacement)
will create CFs
parallel to
this plane.  In some sloshing
scenarios \citep{ascasibar06} the CFs extend considerably above the
plane, but would never cover all
viewing angles; an observed
closed CF ``ring'' would strongly point towards an SICF.  The statistical
properties of a large CF sample could thus be used to distinguish
between the different scenarios. \citet{ghizzardi10} identify 23
apparently relaxed clusters, 10 of which exhibit cold fronts. They find a
{\it perfect} correlation between central entropy levels and cold fronts, with
the 10 lowest central entropy clusters exhibiting cold fronts. This
correlation is also consistent with convection of low entropy gas from the
centre \citep{markevitch07,keshet10}, but would require that all $10$
CFs are viewed from a favorable viewing angle. We offer here an
alternative interpretation, in which the low
entropy and short cooling times imply relatively more active
AGNs and consequently the formation of shocks on shorter periods
\citep{ciotti09}. The SICFs  naturally cover $4\pi$ of the sky,
alleviating the viewing angle statistical requirement.

{\bf Amplitude.} SICFs have distinct entropy and density contrasts
that depend weakly on shock parameters; $q$ is typically larger than
$\sim 1.4$ (assuming $M_1\geq 2$) and is always smaller than
$q_{max}=2.65$. This is consistent with observations.

{\bf Extent.} If cluster oscillations, as described in
\S~\ref{sec:natural_shocks} are  excited, SICF radii
should be approximately logarithmically spaced.
Any shock that expands and collides with the virial shock will create
an SICF at the location of the virial shock, far beyond the core.
The external SICFs are
younger, and would appear sharper owing to less degradation due to the
processes discussed in \S~\ref{sec:stability}. This concentric CF
distribution thus resembles tree rings. Deep observations capable of
detecting CFs at $r\sim 1~{\rm Mpc}$  are predicted to find SICFs
(fig.~\ref{fig:prof}). Such distant CFs occur naturally only in the
SICF model \citep[observations by Suzaku may be able
to probe this range with sufficient quality in the near future;
][]{hoshino10}.

{\bf Plasma diagnostics.}  Shocks are known to modify plasma
properties in a non-linear manner, for example by accelerating
particles to high energies and amplifying/generating magnetic
fields. The plasmas on each side of an SICF may thus differ, being
processed either by two shocks or by one, stronger shock.  This may
allow indirect detection of the CF, in particular if the two shocks
were strong before the collision.  For example, enhanced magnetic
fields below the CF may be observable as excess synchrotron emission
from radio relics that extend across the CF, in nearby clusters, using
future high-resolution radio telescopes (MWA,LOFAR, SKA).

\section{Summary and Discussion}
\label{sec:summary}

Our study consists of two parts. The first is a general, analytic discussion
about cold fronts that form as a result of a merging between a primary and
secondary shock propagating in the same direction. This is a novel
mechanism to create cold fronts discussed here for the first time.
The second part
describes the possible relevance of this SICF mechanism in cluster
environments using 1D spherical hydrodynamical simulations.

We have shown in \S~\ref{sec:riemann} that when shocks moving in the
same direction merge they generate a CF. The density contrast across
the CF is calculated as
a function of the Mach numbers of the two shocks. It is typically
larger than $1.4$ (if $M\gtrsim2$), and is always smaller than
$\myr_{max}=2.65$. We support the analytical calculation by a detailed
investigation of a shock tube planar hydrodynamical simulation.

In \S~\ref{sec:staticdm} and \S~\ref{sec:virial}, using a 1D spherical hydrodynamic code, we demonstrate that SICFs in
clusters are a natural consequence of shocks that are generated at
centres of clusters. We entertain two ways to invoke shocks: by
injecting large amounts of energy near the centre (corresponding to AGN
outbursts), and by abruptly changing potential the well of the cluster
(corresponding to merger events).
 Finally, by simulating cluster evolution from initial cosmological
 perturbations over a Hubble time we show that outgoing shocks that
 merge with the virial shock create very distinctive CFs. These shocks
 can be caused by a critical event (merger or explosion) but can also
 be invoked by stochastic oscillations of the cluster's core, caused
 by  accretion of low angular momentum substructure that perturbs the
 core. We show that a reverberation
mode exists in the haloes of galaxies and clusters that causes
periodic merging between the virial shocks and secondary shocks,
producing an SICF every cycle.
 The simulated SICF contrast is
consistent with the theoretical predictions of \S~\ref{sec:riemann}.
A more thorough investigation of this potentially important mode,
including its stability in 3D is left for future works.

We then discuss in \S~\ref{sec:stability} the
survivability and degradation in time of CFs after they are formed. The
CF discontinuity is smeared over time by diffusion,
at a rate that depends on the unknown nature and amplitude of magnetic
fields.
CFs are susceptible to heat-flux-driven buoyancy instability
\citep[HBI; ]{parrish08},
which could align the magnetic field tangent to the CF and potentially
moderate further diffusion. SICFs, like all other CFs, are subject to
Richtmyer-Meshkov instabilities from subsequent shocks passing through
the cluster. Such collisions also reduce the CF contrast until it reaches
$q\sim 2$.  Unlike most other CF models, an SICF is not expected to
suffer from KHI.

The predicted properties of SICFs are presented in
\S\ref{sec:observe}, and reproduce some of the CF features discussed
in \citet{owers09} and \citet{ghizzardi10}.  In particular, we suggest that CFs in relaxed
clusters, with no evidence of mergers, shear, or chemical
discontinuities, may have formed by shock collisions.  We list the
properties of SICFs that could distinguish them from CFs formed by other
mechanisms.
The SICF model predicts quasi-spherical CFs which are concentric about
the cluster centre, with contrast $\myr \sim 2,$
and possibly extending as far out as the virial shock.
An observed closed (circular/oval) CF could only be an SICF.
In the specific case of cluster reverberation,
a distinct spacing pattern between CFs is expected.
It may be possible to detect them indirectly, for example as
discontinuities superimposed on peripheral radio emission.

Shocks originating from the cluster centre naturally occur in feedback
models that are invoked to solve the overcooling problem.  They are
also formed by mergers of substructures with the BCG.  Thus, SICFs
should be a natural phenomenon in clusters.  Further work is needed to
assess how common SICFs are with respect to other types of CFs, and to
characterize inner SICFs that could result, for example, from
mergers between offset AGN shocks.  The properties of SICFs
in 3D will be pursued in future work.

\section*{Acknowledgements}
We thank I. Parrish, Du{\v s}an Kere{\v s} and M. Markevitch for
useful discussions and the referee, Trevor Ponman, for helpful
suggestions.  YB acknowledges the 
support of an ITC fellowship from the Harvard College Observatory.  UK
acknowledges support by NASA through Einstein Postdoctoral Fellowship
grant number PF8-90059 awarded by the Chandra X-ray Centre, which is
operated by the Smithsonian Astrophysical Observatory for NASA under
contract NAS8-03060.

\bibliography{yuval}

\appendix
\onecolumn
\section{SICF parameters}
\label{sec:gamma53}

Using the thermal jump conditions, we find the density/temperature contrast across the CF given by
\begin{eqnarray}
\myr \equiv \frac{\rho_{3\myin}}{\rho_{3\myout}} =
\frac{M_0^2
  M_1^2(\gamma+1)[M_f^2(\gamma-1)+2]}{M_f^2[M_0^2(\gamma-1)+2][M_1^2(\gamma-1)+2]}
  \left\{ \frac{[(2M_0^2-1)\gamma+1][(2M_1^2-1)\gamma+1]}{(\gamma+1)[(2M_f^2-1)\gamma+1]} \right\}^{-1/\gamma}\, ,
\nonumber
\end{eqnarray}
where $M_f\sim M_0 M_1$ is the Mach number of the newly formed shock. It is determined by including the velocity jump conditions; for $\gamma=5/3$ this yields
\begin{eqnarray}
\left[4\frac{5 M_f^2-1}{(5M_0^2-1)(5M_1^2-1)}\right]^{1/5} = 1
+\frac{M_1^2-1}{\sqrt{(5M_1^2-1)(M_1^2+3)}}
- \frac{4M_1(M_f-M_0)(M_0M_f+1)}{M_f\sqrt{(5M_0^2-1)(M_0^2+3)(5M_1^2-1)(M_1^2+3)}}
\, , \,\,\,\,\,\, 
\end{eqnarray}
which has three roots for $M_f$, only one of which is real.

In the strong primary shock regime $M_0\gg1$, appropriate for example for a secondary
merging with a virial shock, $\myr$ is the root of
\begin{eqnarray} \label{eq:qGeneral}
\myr^{\frac{\gamma-1}{2}} + \myr^{\frac{\gamma}{2}} \sqrt{\frac{\gamma-1}{2\gamma}} = A + O\left(M_0^{-2}\right) \, ,
\end{eqnarray}
where
\begin{eqnarray} \label{eq:AGeneral}
A  \equiv \left( \frac{\gamma+1}{\gamma-1+2M_1^{-2}} \right)^{\frac{\gamma-1}{2}}
 + \left( \frac{\gamma+1}{\gamma-1+2M_1^{-2}} \right)^{\frac{\gamma}{2}}
\frac{\left(M_1-\frac{1}{M_1} \right) \frac{\gamma-1}{\sqrt{\gamma+1}}+ \sqrt{\frac{\gamma^2-1}{2\gamma}}} {\sqrt{(2M_1^2-1)\gamma+1}} \nonumber \, .
\end{eqnarray}
The contrast reaches its maximal value $\myr_{max}$ when $M_1$ satisfies $dA/dM_1=0$. This maximal value is a function of $\gamma$, as shown in Figure \ref{fig:qmax}.

For $\gamma=5/3$, Eqs.~(\ref{eq:qGeneral} simplifies to
\begin{eqnarray}
\left(\frac{M_1^2+3}{4M_1^2}\myr\right)^{1/3} \left(1+\sqrt{\frac{\myr}{5}}\right) =
1-\frac{1-\frac{4M_1}{\sqrt{5}}-M_1^2}{\sqrt{(5M_1^2-1)(M_1^2+3)}} \,. \,\,\,\,\,\,\,
\end{eqnarray}
Maximizing the contrast yields $\myr_{max}=2.653$, corresponding to $M_1=6.654$.

It is interesting to point out that the observed contrast $\myr$ across an SICF could be used to impose an upper limit on $\gamma$, and hence constrain the equation of state.
Figure \ref{fig:qmax} shows that $\myr$ monotonically decreases with increasing $\gamma$ in the relevant, $\gamma>1.15$ regime.
After an SICF is produced with some initial contrast $\myr\leq\myr_{max}$, its contrast typically declines in time as the CF is degraded by diffusion, convection, interactions with subsequent outgoing shocks, etc.
Therefore, $\myr_{max}(\gamma)>\myr$ constrains $\gamma$.
For example, an observed SICF contrast $\myr> 2.653$ ($\myr> 3.108$) would imply that $\gamma< 5/3$ ($\gamma< 4/3$), suggesting the abundance of cosmic rays where the shocks collided.

\label{lastpage}
\end{document}